\documentclass[aip,sd,amsmath,amssymb,reprint,]{revtex4-1}
\usepackage[colorlinks=true,linkcolor=red,urlcolor=black,citecolor=blue]{hyperref}
\usepackage{graphicx}
\usepackage{epstopdf}
\usepackage{dcolumn}
\usepackage{bm}
\usepackage{color, soul}
\usepackage{amssymb}
\begin{document}
\preprint{AIP/123-QED}
\title{Intermittent large deviation of chaotic trajectory in Ikeda map: Signature of extreme events}
	\author{Arnob Ray}
	\affiliation{Physics and Applied Mathematics Unit, Indian Statistical Institute, 203 B. T. Road, Kolkata 700108, India}
	\author{Sarbendu Rakshit}
	\affiliation{Physics and Applied Mathematics Unit, Indian Statistical Institute, 203 B. T. Road, Kolkata 700108, India}
	\author{Dibakar Ghosh}\email{dibakar@isical.ac.in}
	\affiliation{Physics and Applied Mathematics Unit, Indian Statistical Institute, 203 B. T. Road, Kolkata 700108, India}
	\author{Syamal K. Dana}
	\affiliation{Department of Mathematics, Jadavpur University, Kolkata 700032, India}
	\affiliation{Division of Dynamics, Technical University of Lodz, Stefanowskiego 1/15, 90-924 Lodz,Poland}

\begin{abstract}
 We notice signatures of extreme events-like behavior in a laser based Ikeda map. The trajectory of the system occasionally travels a large distance away from the bounded chaotic region, which appears as  intermittent spiking events in the temporal dynamics. The large spiking events satisfy the conditions of extreme events as usually observed in dynamical systems. The probability density function of the large spiking events shows a long-tail distribution consistent with the characteristics of rare events. The inter-event intervals obey a Poisson-like distribution. We locate the parameter regions of extreme events in phase diagrams. Furthermore,  we study two Ikeda maps to explore how and when extreme events terminate via mutual interaction. A pure diffusion of information exchange is unable to terminate extreme events where synchronous occurrence of extreme events is only possible even for large interaction. On the other hand, a threshold-activated coupling can terminate extreme events above a critical value of mutual interaction.
\end{abstract}

\pacs{89.75.Fb, 05.45.−a} 
\maketitle
\begin{quotation}
 Rare and recurrent large amplitude deviations of normally bounded dynamics are seen in many systems. Such occasional large amplitude spiking events are larger than a nominal value and their statistical distribution of occurrence shows qualitative similarities, in dynamical sense, with data records of natural disasters, rogue waves, tsunami, flood and share market crashes. These observations draw attention of researchers to investigate similar sudden large intermittent events in dynamical systems for developing an understanding of the origin of extreme events and exploring the possibilities of prediction. A laser based Ikeda map was studied earlier to profess the origin of a new dynamical phenomenon, namely, interior crisis, that leads to a sudden expansion of a chaotic attractor. This sudden expansion of attractor is not always a permanent property of the system, and it could be intermittent, which shows similarities with extreme events and this signature was overlooked earlier. Here we explore this extreme value dynamical features of the Ikeda map to confirm the phenomenon and the statistical properties of events. An investigation with two coupled maps has also been made in search of an appropriate coupling scheme that is able to terminate these undesirable extreme events.
\end{quotation}

\section{Introduction}
\par  Extraordinary immense events such as rogue waves in the ocean \cite{rough}, harmful algal blooms in marine ecosystems \cite{algal}, epidemics, epileptic seizures \cite{brain}, other natural events such as floods, tsunamis, earthquakes, cyclones, droughts \cite{wether} are extreme events when sudden and rare changes in the nominal behavior are noticed. Man-made systems such as large-scale power blackouts in power supply networks \cite{black}, mass panics \cite{mass}, wars \cite{wars}, share market crashes \cite{market}, regime shifts in ecosystems \cite{regime}, financial crises \cite{risk} are also examples of extreme events. Similar characteristic behaviors are noticed in many dynamical systems where intermittent large deviations in amplitude of a state variable are seen in their temporal dynamics \cite{rajat1,kantz_book}. A sudden large amplitude event is considered as extreme when it deviates from a nominal value by a few standard deviations \cite{cole,th1,th2}. This draws attention of researchers to recreate extreme events in dynamical systems and to develop an understanding of the mechanisms of their origin.

\par  A local instability leads to the origin of this occasional large deviation from the nominal value when the trajectory of a dynamical system arrives near a region of instability in state space\cite{sapsis}. Interior crisis \cite{crisis_pra,grebogi_prl} is identified as one  possible reason for the origin of extreme events in many dynamical models \cite{rajat2,th2,super, kingston}, experimental systems such as laser systems \cite{optics} and electronic circuits \cite{noise,circuit}. A chaotic attractor shows sudden, but occasional and recurrent large expansion in size that is manifested as intermittent bursting out of the phase-space region within which the attractor was originally confined before a crisis. Such an abrupt transition also occurs in dynamical systems where a laminar phase (or almost periodic oscillation) encounters intermittent transition to a turbulent flow or a chaotic flow of larger amplitude \cite{intermittency, Dana}. In a microelectro-mechanical system, such rare transitions are also found to occur between coexisting orbits via sliding bifurcation \cite{mems}. In multistable systems, on the other hand, a sudden transition from one state to another causes an extreme event under the influence of noise \cite{noise}. Noise is responsible for intermittent switching between the coexisting states \cite{noise,switch}.  Manifestation of extreme events has also been reported in coupled systems. The trajectory of the coupled systems, after realization of synchrony above a critical coupling, mos of the time lies on a synchronization manifold. This trajectory occasionally travels away from the synchronization manifold along the transverse direction due to local instability originated by the presence of noise or parameter mismatch. This is called attractor bubbling, appearing via bubbling transition \cite{ott,ashwin}, and blow-out bifurcation \cite{arindam,rajat1,instab} leading to the origin of occasional extraordinary large events, which are classified as extreme events \cite{predict,Dana,arindam}.

\par  Besides originating extreme events-like behavior in dynamical systems, an urgent task is prediction of extreme events, althoigh it is very difficult. So far some attempts have been made in search of early warning of extreme events \cite{predi,farazmand}, but with not so much success. Even in deterministic dynamical systems, this is a challenging issue of current research \cite {predict,sapsis,jordi}. Understanding the origin of this complex dynamical process is a first priority before prescribing any algorithm for prediction of extreme events.

\par  In this paper, a laser based Ikeda map is considered for our study where a sudden large expansion of a chaotic attractor via interior crisis has been reported earlier \cite{grebogi_prl}. This sudden transition of the attractor may not always lead to a permanently enhanced attractor of larger size. Sudden expansion of the attractor could be intermittent, which we claim as showing signatures of extreme events as supported by statistical properties of long-tail probability distribution of events. The parameter regions of extreme events are located in the Ikeda map. Furthermore, we study two coupled Ikeda maps, first using a simple diffusive coupling and then a threshold controlled coupling, which provide  interesting information how we can trigger and terminate extreme events in the coupled system. It is shown that a purely diffusive mutual interaction between two maps never could terminate, but produce only synchronous extreme events-like intermittent spiking.  On the other hand, a threshold-activated coupling is able to suppress extreme events for a range of mutual interactions, although, events are occurring in an uncorrelated manner in the coupled systems.

\section{Model}\label{model}
\par  We consider a simplified version \cite{ikeda2} of the Ikeda map
\begin{equation}\label{eq:system1}\small
\begin{array}{lcl}
z_{n+1}=A+B~z_n\exp\Big[ik-\frac{ip}{1+|z_n|^2}\Big],
\end{array}
\end{equation}
where $z_n=x_n+iy_n$ $\big(i=\sqrt{-1}\big)$. This model describes the evolution of laser across a nonlinear optical resonator. The real valued two dimensional Ikeda map is derived as
\begin{equation}\label{eq:system}\small
\begin{array}{lcl}
x_{n+1}=A+Bx_{n}\cos\Big(k-\frac{p}{w_n}\Big)-By_{n}\sin\Big(k-\frac{p}{w_n}\Big),\\[5pt]
y_{n+1}=By_{n}\cos\Big(k-\frac{p}{w_n}\Big)+Bx_{n}\sin\Big(k-\frac{p}{w_n}\Big),
\end{array}
\end{equation}
where $A$ is the laser input amplitude, $B$ is the coefficient of reflectivity of the partially reflecting mirrors of the cavity, $w_n={1+x_{n}^2+y_{n}^2}$ and $k$ is the laser-empty-cavity detuning, and $p$ measures the detuning due to the presence of a nonlinear medium in the cavity\cite{des}. We kept $A=0.85, B=0.9, k=0.4$ fixed all throughout the text and consider $p$ as a control parameter, when we are able to originate extreme events-like spiking behavior.

\section{Results}\label{results}
\par
\begin{figure}[ht]
	\centerline{\includegraphics[scale=0.5]{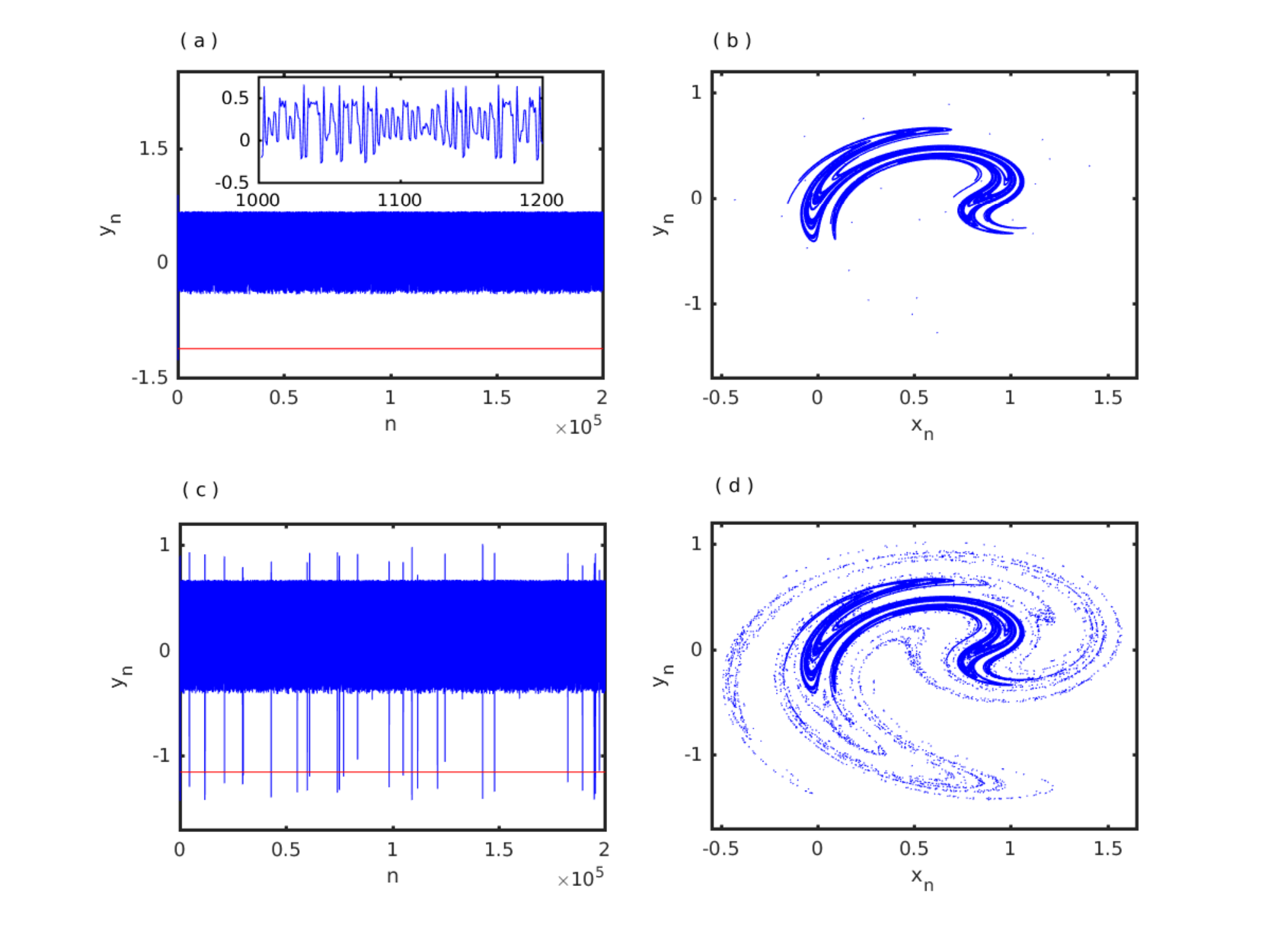}}
	\centerline{\includegraphics[scale=0.5]{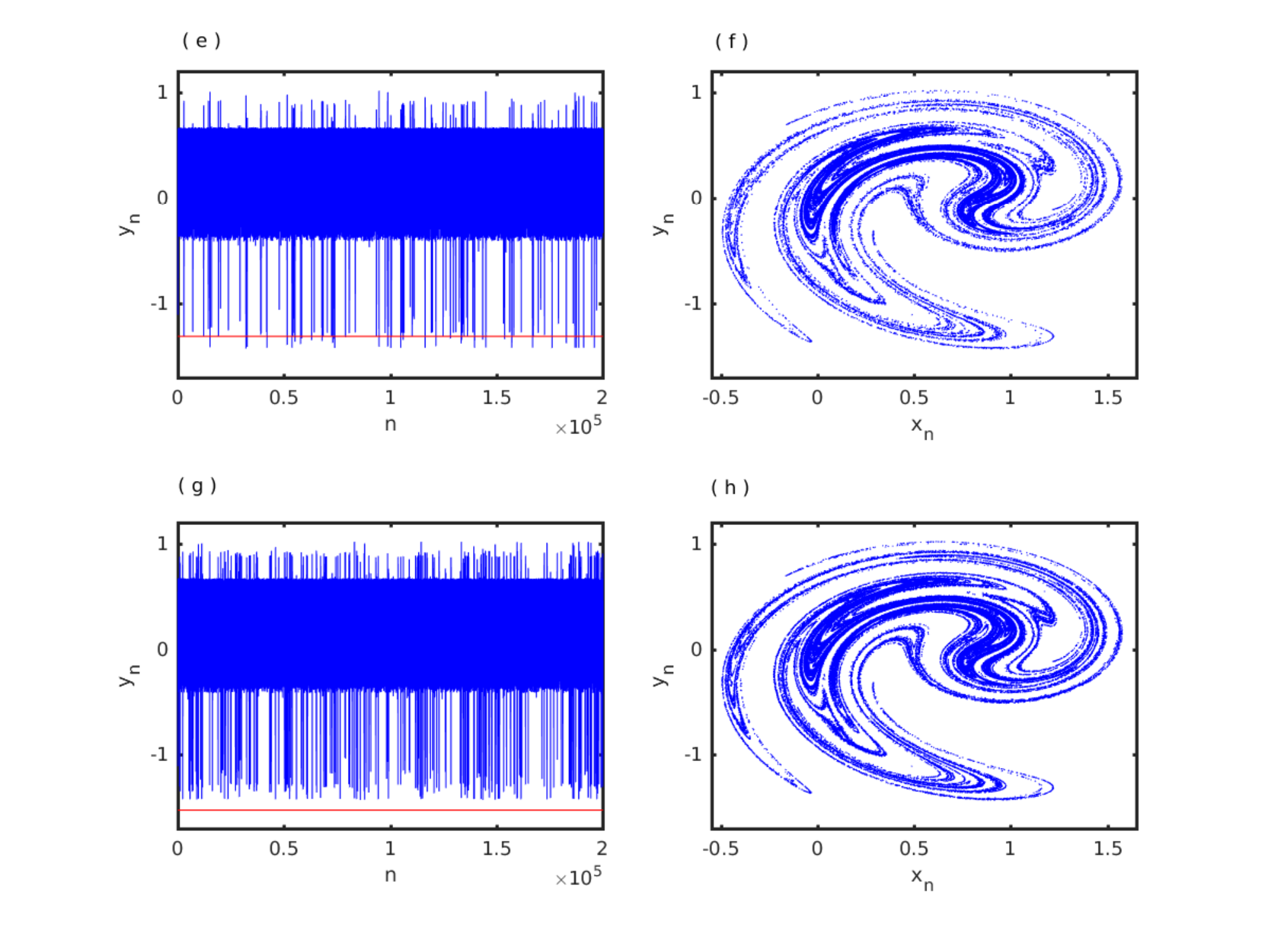}}
\caption{Temporal dynamics (left panels) and phase portraits (right panels) of Ikeda map. (a, b) pre-crisis bounded chaos for $p=7.265$, inset of (a) represents the temporal dynamics for a short time interval, (c, d) post-crisis extreme events for $p=7.275$, (e, f) extreme events (more frequent) for $p=7.285$, (g, h) very frequent large events for $p=7.295$. Horizontal lines (red) in left panels indicate extreme event qualifier threshold $T$.}
	\label{fig1}
\end{figure}

 We take a long run $(1.0\times10^{10}~\text{iterations})$ of $y_n$, measure the local minimum values, $P_n$= min($y_n$), and estimate their mean $\mu=\langle P_n \rangle$ and standard deviation $\sigma=\langle P_n^2 \rangle-\mu^2$. An event is then classified as extreme when it crosses a threshold, $T=\mu-d\sigma$. Noteworthy that there is no strict quantitative definition of extreme events so far, especially, in natural surroundings; even a smaller event can make huge damage of infrastructure and life. An arbitrary threshold limit of $d~(4 \text{ to } 8)$, that indicates a few standard deviation away from the nominal value of a time series or temporal evolution of a state varible, is effectively used to classify extreme events in many dynamical systems \cite{dysthe,th1,th2,Dana}. We choose $d=5$, in our work, and extreme events are lower than the value $T$ since the events are negative-valued. To compute this threshold value a long run of iteration is taken until $T$ becomes saturated. Temporal dynamics and phase portraits of the iterated Ikeda map are shown in Fig.~\ref{fig1} with changing $p$. Left panels show temporal behaviors of $y_n$, and the corresponding phase portraits ($x_n$ vs. $y_n$) are plotted in the right panels. Figure \ref{fig1}(a) shows temporal evolution of {\it pre-crisis} bounded chaos for $p=7.265$. In the inset figure, we have plotted the time series for a short time interval to show the variation of local extrema of $y_n$. No large spike or burst is observed here, accordingly, its trajectory is shown bounded in a dense region of phase-space in Fig.~\ref{fig1}(b). Occasional large amplitude spikes are seen in Fig.~\ref{fig1}(c) for a larger value of $p=7.275$. The phase portrait in Fig.~\ref{fig1}(d) reveals a dense blue region, but it shows an extended region of sparsely distributed points (blue dots) representing the occasional large spikes as shown in the temporal dynamics in Fig.\ref{fig1}(c). Their inter-event return times are irregular and distinctly large, which signify their rare occurrence. We classify those intermittent large events as extreme by applying the threshold measure $T$ on the state variable $y_n$; horizontal red lines represents the qualifier threshold $T$ in the left panels. For a larger $p=7.285$, more frequent large amplitude events are seen in Fig.~\ref{fig1}(e). Accordingly, the outer periphery of the phase portrait in Fig.~\ref{fig1}(f) no more remains sparse compared to Fig.~\ref{fig1}(d). For a larger $p=7.295$, very frequent spikes is seen in Fig.~\ref{fig1}(g), but no large spike exceeds the threshold $T$ and thereby fail to qualify as extreme events. The frequent large spikes increase the mean value $\mu$, which causes to lower the $T$ value. Figure~\ref{fig1}(h) is dense now; the trajectory now travels the enlarged phase space very frequently, basically enlarging the bounded chaos in size permanently (cf. Fig.~\ref{fig1}(b) and Fig.~\ref{fig1}(h)).

\par
\begin{figure}[ht]
	\centerline{\includegraphics[scale=0.488]{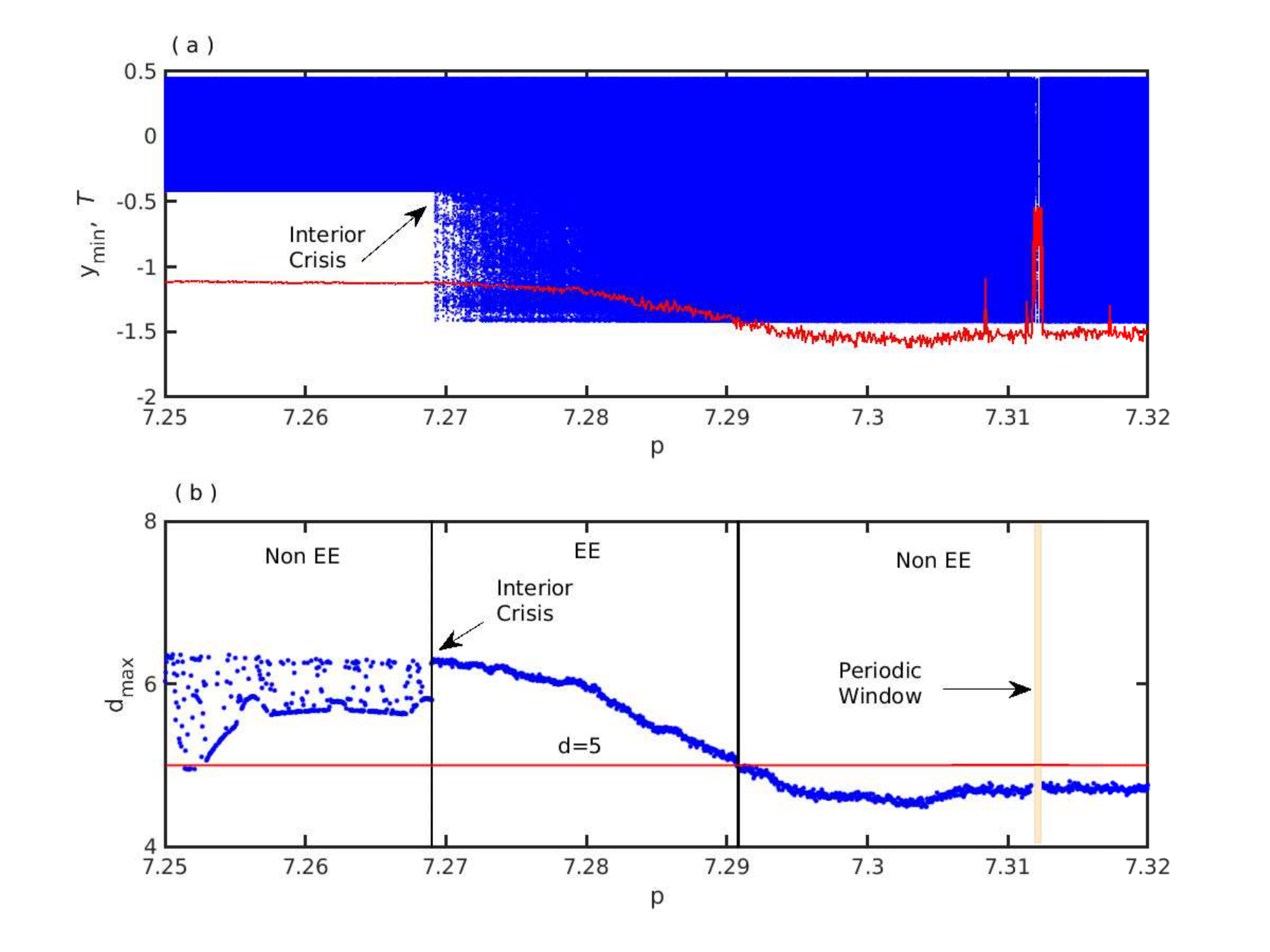}}
	\caption{(a) Bifurcation diagram of $y_{min}$ (blue dots) and extreme event qualifier threshold $T$ (red curves) against $p$. Sudden expansion of attractor at a critical point $p_{c}=7.26884894$ via interior crisis. (b) Variation of $d_{max}$ (blue dots) with $p$. Red horizontal line indicates, $d=5$ line,  extreme events region is marked between black vertical lines.}
	\label{fig2}
\end{figure}

 A bifurcation diagram is plotted in Fig.~\ref{fig2}(a)  to visualize the changing scenario with $p$ that indicates a critical point $p=p_c$ (interior crisis, indicated by a black arrow) where the size of the attractor abruptly expands from a pre-crisis bounded chaos. At the crisis point, $y_{min}$ becomes suddenly large when the trajectory of the system starts occasional travel to a larger phase space away from the bounded dense region as shown in  Fig.\ref{fig1}(d).  For a gradual increase of $p$ beyond this crisis point, the attractor remains larger due to occasional  far away travel, however, $y_{min}$ becomes more dense. The threshold $T$ plot (red line) recognizes emergence of extreme events beyond the crisis point until $p=7.29$. This corroboartes the {\it pre-crisis} and {\it post-crisis} scenarios in phase space and their temporal dynamics in Fig.~\ref{fig1}.  We do not repeat here how interior crisis emerges in Ikeda map since it has already been established earlier \cite{crisis_pra, grebogi_prl} for this system. Our focus is rather on evidence of extreme events, and statistical distribution of events and inter-event intervals (IEI), which  are not yet explored, to the best of our knowledge.

\par For a confirmation of the critical point $p=p_c$, a measure $d_{\max}$ is defined \cite{super,mems}
\begin{equation}\label{eq:n}
\begin{array}{lcl}
d_{\max}=\dfrac{\mu-\min{(y_{n})}}{\sigma},
\end{array}
\end{equation}
where $\min(y_n)$ is a minimum value of $y_n$ observed in a long-time series. Figure~\ref{fig2}(b) plots $d_{\max}$ with varying $p$ that clearly identifies a critical point when the sudden expansion of attractor occurs that is expressed as occasional large spiking events to turn on. It shows strong fluctuation in the pre-crisis region, which stop fluctuating and a monotonic decreasing trend follows in the post-crisis region due to increasingly more frequent occurrence of large spiking. The crisis point $p=p_c$ is almost exactly identified from this $d_{max}$ plot. A window of $p$ interval is noticed where extreme events continue to emerge in the system, and it matches with the bifurcation plot of $y_{min}$ above. It starts at $p_c$ where crisis starts and turns off at a higher $p$ where the $d_{max}$ curve intersects the horizontal line $d=d_{max}=5$, which is used to define our extreme value threshold, $T=\mu-d \sigma$. Beyond this $p$ window, $d_{max}$ value is almost saturated when spiking becomes more frequent and lost the character of extreme events.

\begin{figure}[ht]
	\centerline{\includegraphics[scale=0.4]{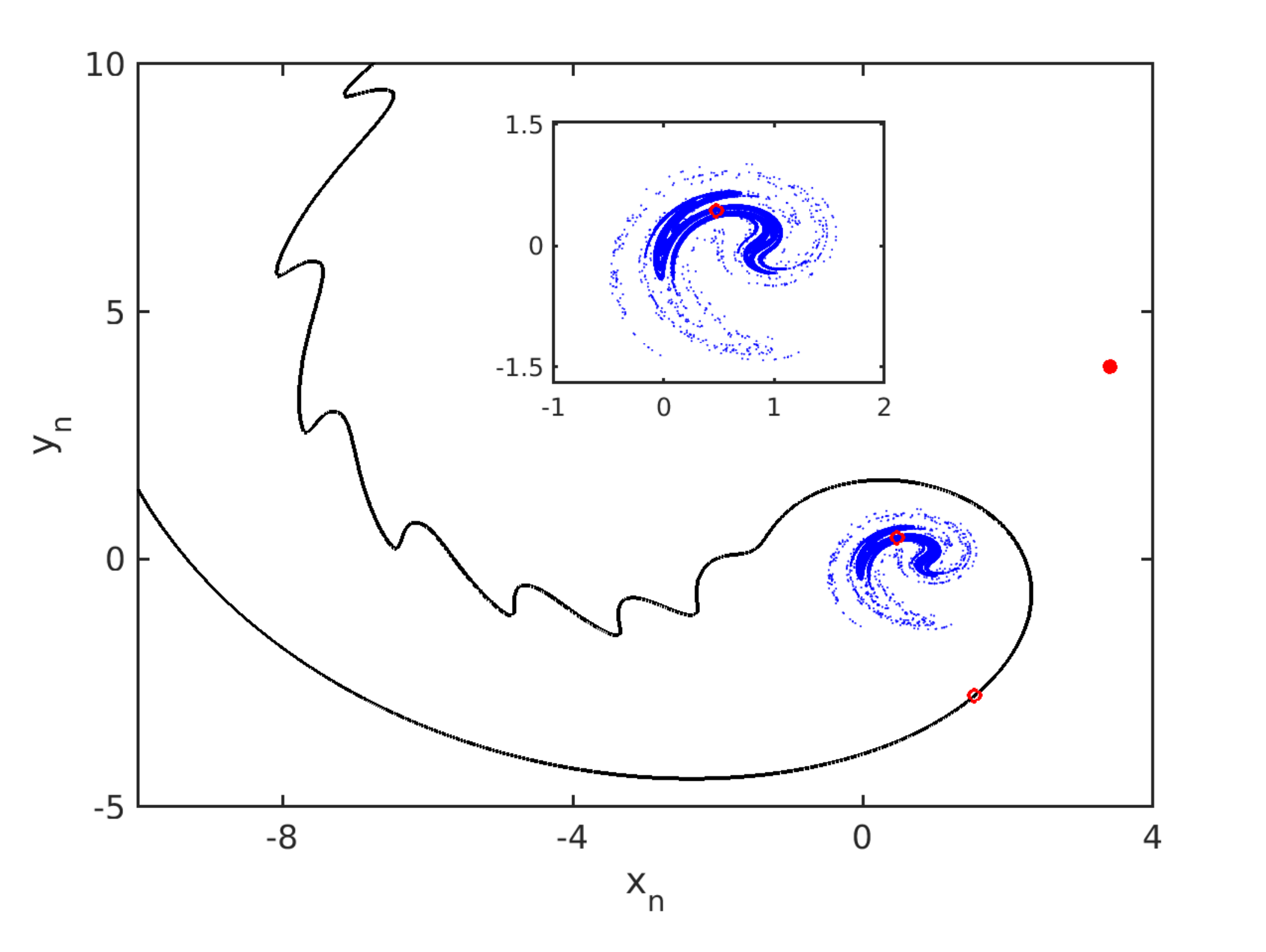}}
\caption{Basin of attraction of the Ikeda map. A boundary line (black) separates the basin of the chaotic attractor  and the coexisting stable fixed point (solid red circle) for $p=7.275$. One unstable fixed point (open red circle) lies on the basin boundary and another unstable fixed point lies on the chaotic attractor (blue) inside the boundary (black line). Inset shows an enlarged version of the attractor during extreme events.}
\label{fig3}
\end{figure}
\par  The system has one stable fixed point  with two coexisting unstable fixed points. The basin of attraction of the system is plotted in Fig.~\ref{fig3} for a post-crisis parameter value $p=7.275$. The basin of the chaotic attractor (blue) is delineated by a boundary line (black line). An enlarged view of the post-crisis attractor is shown in the inset. One stable, two saddle points are denoted by a solid circle, open circles, respectively. One saddle point $(0.4712,0.4375)$ is situated on a site close to the chaotic attractor around which it evolves, which introduces the interior crisis \cite{crisis_pra}. Another saddle point $(1.537,-2.754)$ is located on this basin boundary, and this basin boundary is the stable manifold of that saddle point. We have checked that the basin boundary of the system remains unchanged for other values of $p=7.265$ (bounded chaos), $p=7.285$ (frequent extreme event) and $p=7.295$ (non-extreme event).

\begin{figure}[ht]	\centerline{\includegraphics[scale=0.335]{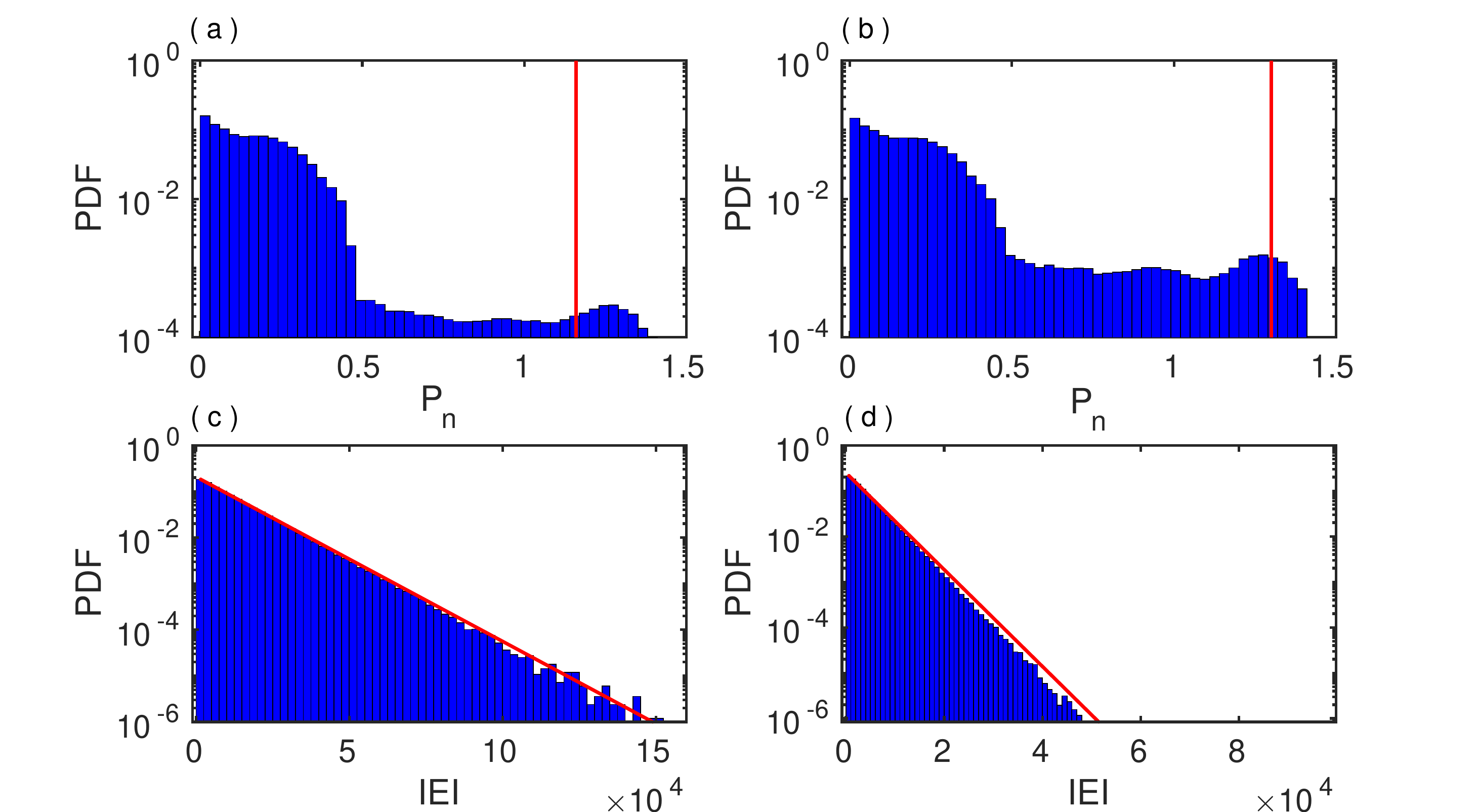}}
\caption{Probability density function of event heights and inter-event intervals. (a,b) Extreme events occur in the tail, vertical red lines indicate the threshold height. (c,d) Probability density function of IEIs, with fitted Poisson distribution by red lines. Parameters for (a,c) $p=7.275$ and (b,d) $p=7.285$. Number of iterations are taken as $1.0\times10^{10}$.}
\label{fig4}
\end{figure}
\par The statistical properties of events are presented in Fig.~\ref{fig4} for two selected values of the parameter $p~(=7.275$ and $7.285)$ from the post-crisis regime. The probability density functions (PDF) of event heights $P_n$ are shown where long-tail distribution of events \cite{long} are noticed in both the cases. The vertical red lines represent our pre-assigned critical threshold $T$, which shifts to a larger $P_n$ value for the second case $(p= 7.285)$, when event spikes becomes more frequent. Figures \ref{fig4}(a) and \ref{fig4}(b) are the PDF of the event height corresponding to the time series presented in Figs.~\ref{fig1}(c) and \ref{fig1}(e), respectively. Larger frequency of occurrence of events are reflected in the thickness of the tail in the latter case of $p=7.285$. The probability density function of inter-event intervals (IEI) are plotted in Figs.~\ref{fig4}(c) and \ref{fig4}(d) for our two example cases. These PDFs are then fitted by $P(r)=\lambda~\mbox{e}^{-\lambda r}$ since IEIs are almost uncorrelated (checked separately, but not presented here) and hence the PDF of IEI follows a Poisson-like distribution \cite{geo}, where $r$ is the inter-event interval and $\lambda (>0)$ is the shape parameter. The estimated values of the shape parameter are $\lambda=0.00008214$ and $\lambda=0.0002448$ for $p=7.275$ and $p=7.285$, respectively. The corresponding Poisson distributions are fitted with their respective distributions of the IEI by red lines. The slope $\lambda$ increases with $p$ and it is consistent with our observation of temporal dynamics that number of events are more frequent with increasing $p$.

\par
\begin{figure}[ht]
	\centerline{\includegraphics[scale=0.37]{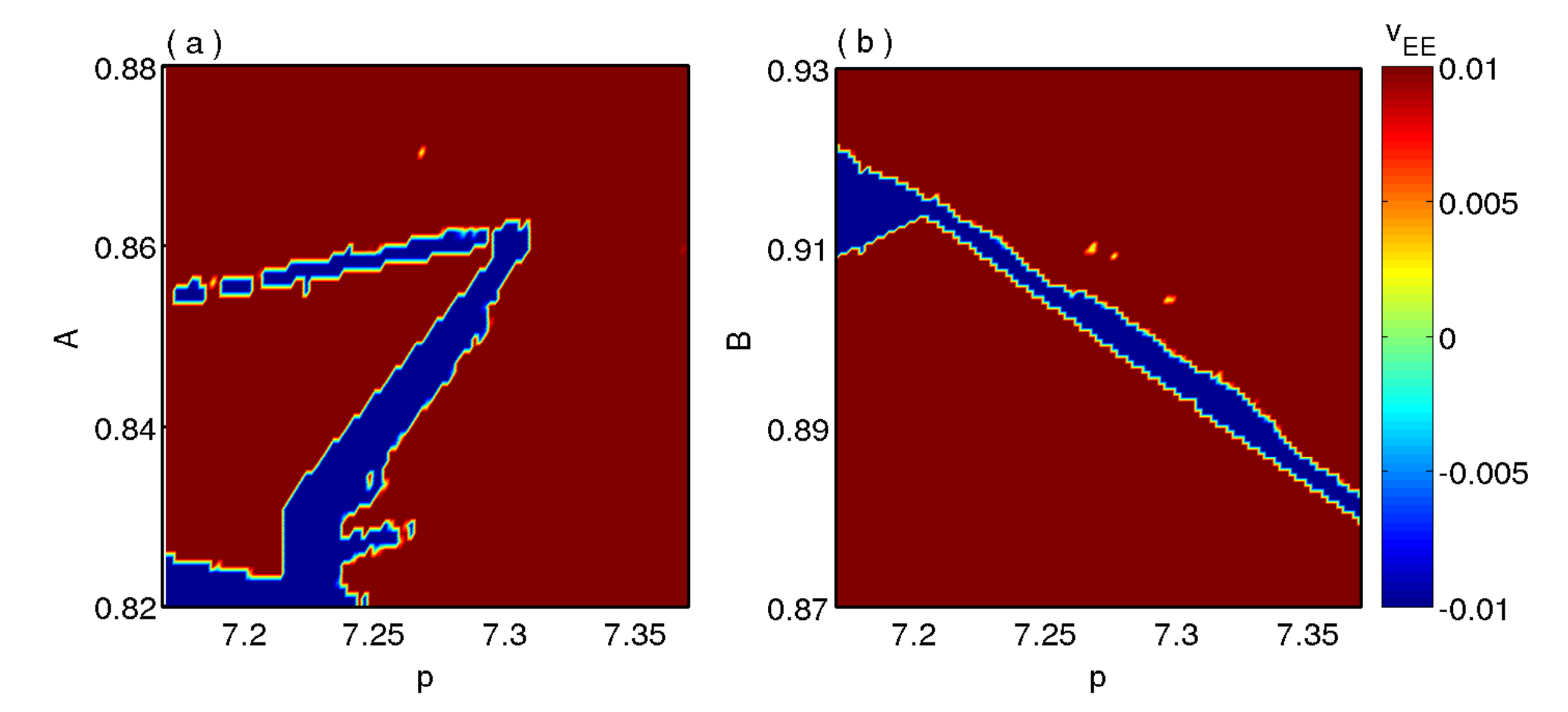}}
	\caption{Phase diagram of extreme events. (a) $(p,A)$ parameter space for $B=0.9, k=0.4$, (b) $(p,B)$ parameter space for $A=0.85, k=0.4$. Color bar represents the value of $v_{EE}$.}
	\label{fig6}
\end{figure}

 The role of other system parameters on the origin of extreme events are also checked. Here we scan the parameter space for extreme events taking two sets of parameters, $(p,A)$, $(p,B)$, separately. To delineate the regions of extreme events and the non-extreme events, in parameter space, a measure $v_{EE}=y_{min}-T$ is defined, where $y_{min}=\min_{n \in \mathbb{Z}}\{y_n\}$. As discussed earlier, an event is defined as extreme if it exceeds the threshold $T$, and in our case, it is when  $y_{min}<T$. For no-extreme events, in an entire time series, not a single spike crosses the threshold, {\it i.e.},  when $y_n \ge T$ is maintained. Therefore $v_{EE}$ will be negative if the event occurs at least once in an entire time series, and positive otherwise. The extreme events and no-extreme events regions are marked by color bar using this measure. The blue regions correspond to the extreme event. Two phase diagrams in Fig.~\ref{fig6} convincingly prove that extreme events are not restricted to our selected parameters, but exists in a reasonably large parameter range of $(p,A)$ and $(p,B)$. The parameter region of $(p,k)$ is also checked that produces similar results, but not presented here.

\section{Coupled Ikeda map}\label{control}
 Next, we address a relevant question if it is possible to continue with extreme-events-like rare and recurrent large spiking events in two or more mutually interacting Ikeda maps. It may provide a clue how mobilites or a diffusion of resources may influence the extent of extreme events. Is it possible to terminate such large events by mutual exchange of information or dispersal related diffuion between patches, say, in an ecological setting? We consider two maps and first use a diffusive coupling, 
\begin{equation}\label{Eq:coupled_2}
\begin{array}{lcl}		
x_{1,n+1}=f(x_{1,n},y_{1,n}),\\[5pt]
y_{1,n+1}=g(x_{1,n},y_{1,n})+\epsilon(g(x_{2,n},y_{2,n})-g(x_{1,n},y_{1,n})),\\[7pt]

x_{2,n+1}=f(x_{2,n},y_{2,n}),\\[5pt]
y_{2,n+1}=g(x_{2,n},y_{2,n})+\epsilon(g(x_{1,n},y_{1,n})-g(x_{2,n},y_{2,n}))
\end{array}
\end{equation}
where $n$ is an integer, $\epsilon$ be the interaction strength,
\begin{equation*}
	\begin{array}{lcl}
		f(x_n,y_n)=A+Bx_n\cos\Big(k-\frac{p}{w_n}\Big)-By_n\sin\Big(k-\frac{p}{w_n}\Big),\\[5pt]
		g(x_n,y_n)=By_n\cos\Big(k-\frac{p}{w_n}\Big)+Bx_n\sin\Big(k-\frac{p}{w_n}\Big).
	\end{array}
\end{equation*}

\begin{figure}[ht]
	\centerline{\includegraphics[scale=0.34]{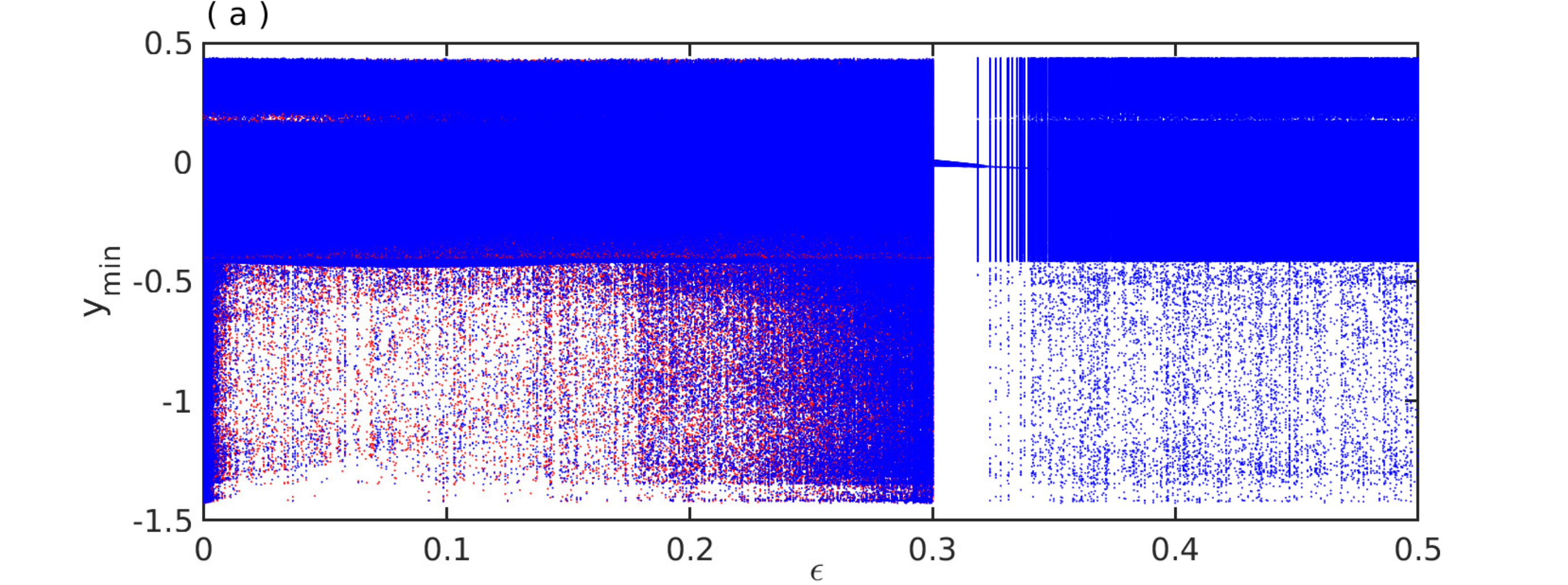}}
\vspace{1pt}
	\centerline{\includegraphics[scale=0.305]{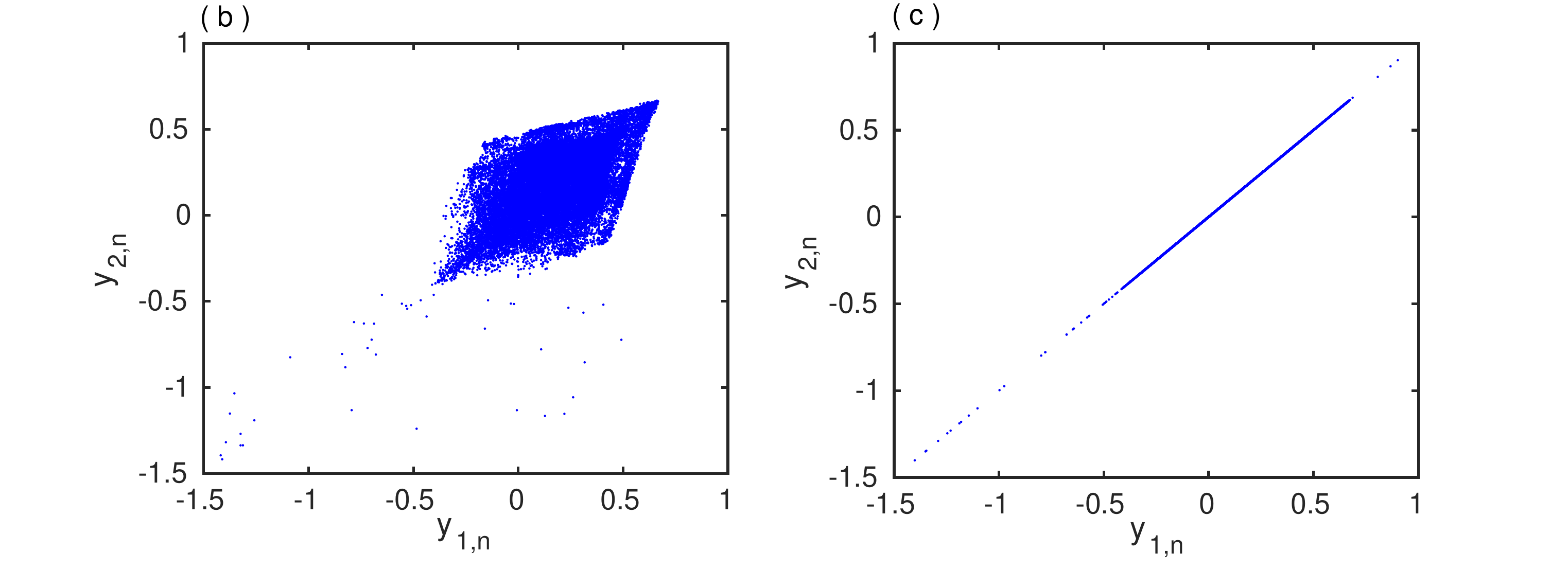}}
	\caption{Diffusive coupling: (a) Bifurcation of temporal evolution of coupled maps with respect to the diffusion strength $\epsilon$. Blue dots: ${y_1}_{min}$, red dots: ${y_2}_{min}$. Plots of synchronization manifold $y_{1,n}$ vs. $y_{2,n}$ plane for (b) $\epsilon=0.2$ and (c) $\epsilon=0.4$. Other Parameters : $p=7.275, A=0.85, B=0.9,$ and $k=0.4$.}
	\label{fig7_1}
\end{figure}

\par We consider a set of identical parameters for both the maps, each of them put into the post-crisis regime, in isolation. A bifurcation diagram ${y_{1,2}}_{min}$, the local minima (blue and red dots) of two system variables, $y_{1,n}$ and $y_{2,n}$  are plotted in Fig.~\ref{fig7_1}(a) with varying $\epsilon$. It clearly shows that two maps continue with extreme events in an incoherent manner in the range of small mutual interactions. Large events appear in both the systems incoherently (blue and red dots). Figure~\ref{fig7_1}(b) verifies the desynchronised events, where a  $y_{1,n}$ vs. $y_{2,n}$ plot indicates the synchronization manifold for a coupling strength $\epsilon=0.2$. Both the systems suddenly turns off all large events at a critical strength $\epsilon=0.3$. Here a large shrinking in the size of the attractors is seen  undergoing a reverse process to periodic motion. We do not focus on this non-event parameter window, at this point, which is not our interest of study now. However, for a further increase of $\epsilon$, extreme events returns suddenly, when both the systems generate such events in a synchronous manner. Coherence or synchrony of events is verified in Fig.~\ref{fig7_1}(c), by another  $y_{1,n}$ vs. $y_{2,n}$ plot for $\epsilon=0.4$.

\par Alternatively, if we introduce a bidirectional threshold-activated-coupling \cite{sinha}, we do not find synchronous extreme events in the coupled maps. Hwever,  this coupling introduces a smooth annihilation of extreme events. If $y_{i,n}$ ($i^{th}$ oscillator) exceeds a critical value $y_c$, a feedback information is transferred to its neighboring $j^{th}$ oscillator with a scaling factor $\epsilon$, {\it i.e.}
\begin{equation}\label{eq:control}
\begin{array}{lcl}
y_{i,n}\rightarrow y_{i,n}-\epsilon(y_{i,n}-y_c)\\[5pt]
y_{j,n}\rightarrow y_{j,n}+\epsilon(y_{i,n}-y_c),
\end{array}
\end{equation}
Using this threshold-activated-coupling, the coupled system is defined as
	\begin{equation}\label{Eq:coupled}
	\begin{array}{lcl}
	x_{1,n+1}=f(x_{1,n},y_{1,n}),\\[5pt]
	y_{1,n+1}=g(x_{1,n},y_{1,n})+\epsilon\theta(y_c-y_{1,n})(y_c-y_{1,n})\\\hspace{94pt}-~\epsilon\theta(y_c-y_{2,n})(y_c-y_{2,n}),\\[7pt]
	x_{2,n+1}=f(x_{2,n},y_{2,n}),\\[5pt]
	y_{2,n+1}=g(x_{2,n},y_{2,n})+\epsilon\theta(y_c-y_{2,n})(y_c-y_{2,n})\\\hspace{94pt}-~\epsilon\theta(y_c-y_{1,n})(y_c-y_{1,n}),
	\end{array}
	\end{equation}
where $\epsilon$ is the strength of interaction or a scaling factor, as usual, $\theta(x)=1$ if $x>0$ and $0$ otherwise.

\begin{figure}[ht]
	\centerline{\includegraphics[scale=0.3]{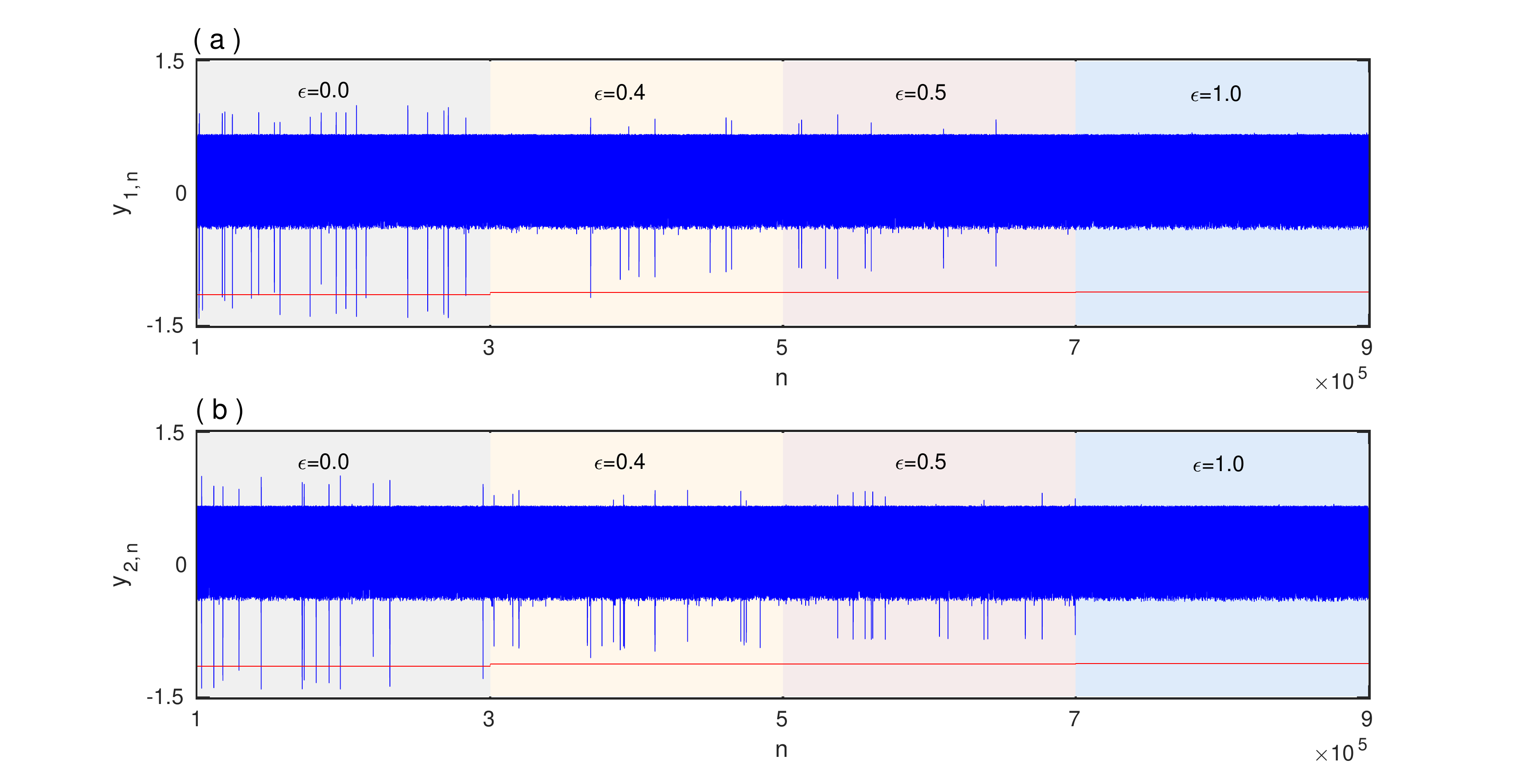}}
\caption{Threshold-activated coupling: temporal evolution of coupled Ikeda maps. (a) $y_{1, n}$ and (b) $y_{2, n}$ for four different values of $\epsilon$. Red horizontal lines represent the qualifier threshold $T$. Other parameters: $p=7.275, A=0.85, B=0.9,$ and $k=0.4$.}
	\label{fig7}
\end{figure}
\par The sequences of temporal dynamics in two Ikeda maps are shown in Figs.~\ref{fig7}(a)-(b) for changing $\epsilon$ values. Short runs of time series for four different $\epsilon$ values are shown in separate colors. Time evolution exhibits occasional large deviation as long as the two sub-systems are isolated. At first, time series of $y_{1, n}$ and $y_{2, n}$ are plotted for $\epsilon=0$ to demonstrate extreme events in isolation. When the coupling is switched-on, the large deviation slowly decays towards it's bounded behavior with increasing $\epsilon$. At $n=3\times10^5$ when $\epsilon=0.4$, the extreme events are almost suppressed in subsystem-2 (Fig.\ref{fig7}(b)), but subsystem-1 (Fig.~\ref{fig7}(a)) till continues with large events, when at least one large spike crosses the threshold height $T$. By turning on the coupling with a larger strength $\epsilon=0.5$ at time $n=5\times10^5$, spiking events continues, but never cross the threshold. Finally, all events are terminated restoring the bounded chaos for large interaction $\epsilon=1.0$. Contrary to purely diffusive coupling, the threshold-activated-coupling is able to suppress extreme events above a critical value of $\epsilon=\epsilon_c$ and furthermore, no coherence of extreme events is seen.

\begin{figure}[ht]
	\centerline{\includegraphics[scale=0.37]{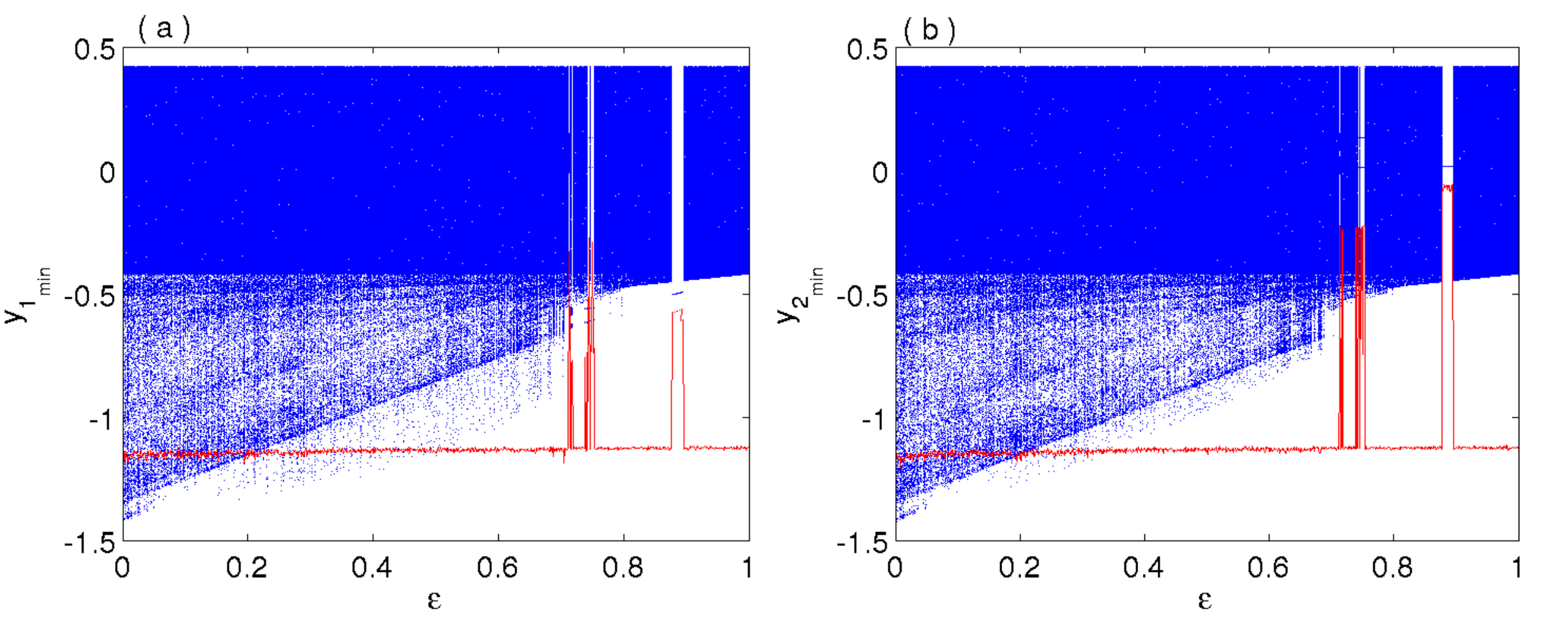}}
	\caption{Bifurcation diagram of the coupled system with respect to threshold-activated coupling strength $\epsilon$ for the variables (a) $y_{1, n}$ and (b) $y_{2, n}$. Red curves be the extreme events qualifier threshold $T$ at the respective values of $\epsilon$. The other parameter values are same as in Fig. \ref{fig7}.}
	\label{fig8}
\end{figure}
To make an impression about the critical point  of coupling strength $\epsilon_c$ where extreme events are terminated, two bifurcation diagrams are plotted for $y_{1, min}$ and $y_{2, min}$ in Fig.~\ref{fig8} for varying $\epsilon$. The threshold $T$ for both the state variables $y_{1, min}$ and $y_{2, min}$ are added with the plots in red curves. The choice of parameters of isolated maps are set for the generation of extreme events as shown above. With a gradual increase of $\epsilon$, the local minima of peaks gradually increases. Above a critical value of $\epsilon_c$, all the negative spikes $P_n$ are now confined above the $T$ line, and extreme events are terminated there. Finally, beyond a larger $\epsilon$ value, the coupled system generates only bounded chaos.

\begin{figure}[ht]
	\centerline{\includegraphics[scale=0.45]{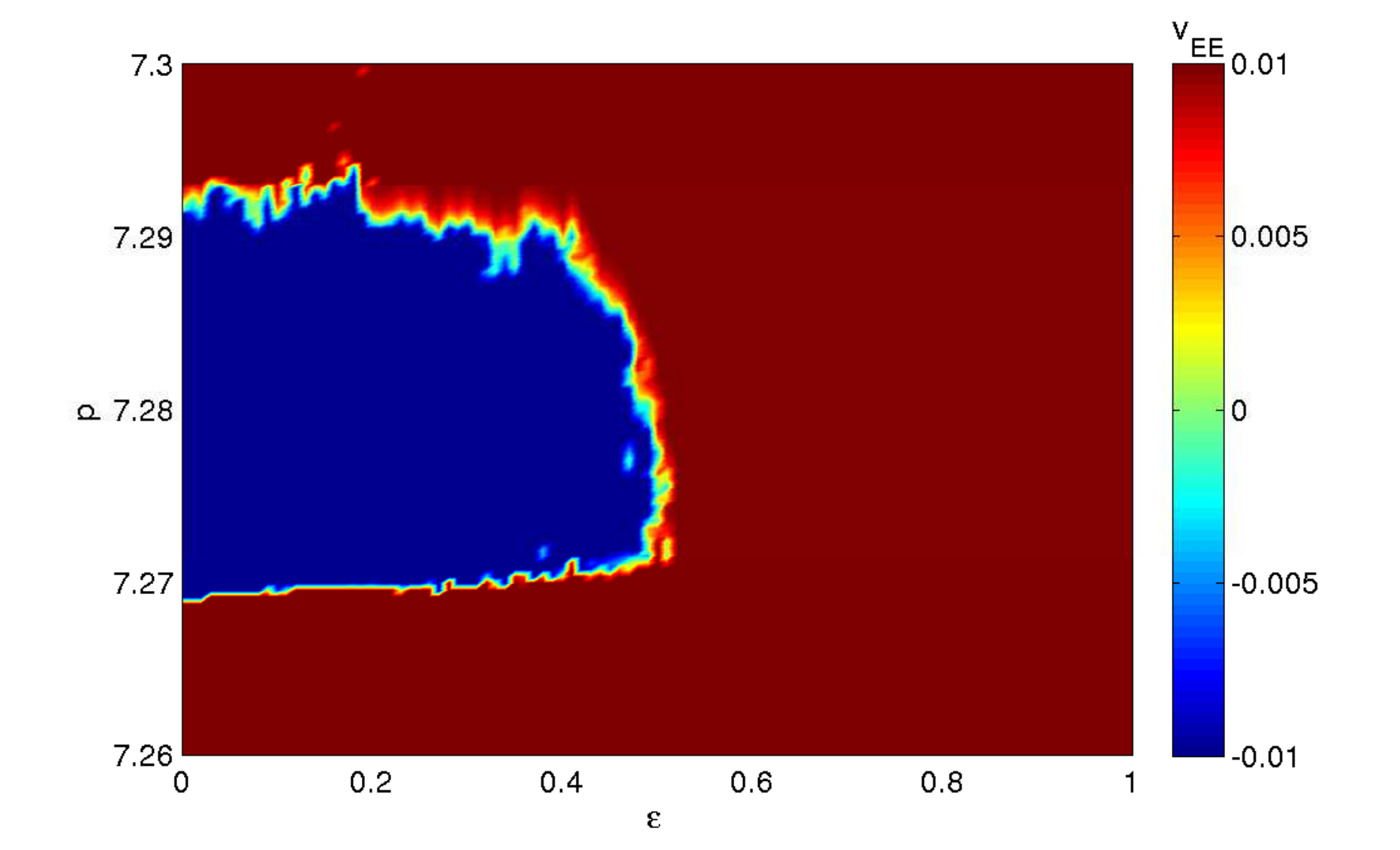}}
	\caption{Phase diagram in $(\epsilon, p)$ parameter plane for occurrence of extreme events (blue region) and non-extreme events (red region) using threshold-activated coupling. The variation of $v_{EE}$ is shown in color bar.}
	\label{fig9}
\end{figure}
A phase diagram in Fig.~\ref{fig9}  represents an overview of the triggering and termination of extreme events in a $\epsilon-p$ plane. $\epsilon$ is varied in the range $[0,1]$ and the system parameter $p$ taken in a range $[7.26,7.3]$. To distinguish the regions of extreme events and no-extreme events, we use the $v_{EE}$ condition once again as discussed above. We notice that for all values of $p\in[7.269,7.291]$ for which the uncoupled system ($\epsilon=0$) exhibits extreme events, there by increasing the coupled strength to $\epsilon$ it can be suppressed above a critical value.

\section{Conclusion}\label{conclusion}
\par A laser based Ikeda map was shown earlier that it encounters an interior crisis leading to a large expansion of a chaotic attractor. It was overlooked then that such large expansion of the attractor could be either an intermittent or a permanent phenomenon. In particular, we focus on intermittent expansion of the attractor at a post-crisis parameter regine and showed that this phenomenon carries the signatures of extreme events as usually elaborated in current literature. The pre-crisis and post-crisis regions in the parameter space are separated where the extreme events emerged and terminated. A threshold size limit of events is defined to assign extreme events and delineate the parameter regions of existence of extreme events and showed that it prevailed in a reasonably large parameter space of the relevant system parameters. The statistical properties of extreme events were studied that confirmed the signatures of the haunted phenomenon. PDF of events heights for two selective post-crisis parameters show long-tailed distribution, which  confimed rare and recurrent nature of typical extreme events in dynamical systems. PDFs of IEI are also plotted for two concerned parameters, which follow Poisson-like distribution. We extended the work to two mutually interacting Ikeda maps. It was shown that a purely diffusive interaction failed to turn off extreme events for a large range of interaction, however, it generated synchronous occurrence extreme events. Alternatively, a threshold-activated-coupling generated only asynchronous extreme events and smoothly suppresssed the extreme events with increasing interaction between the maps.\\

{\bf Acknowledgments:}
D.G. was supported by SERB-DST (Department of Science and Technology), Government of India (Project no. EMR/2016/001039). S.K.D. is supported by UGC (India) Emeritus fellowship scheme no. Emeritus-2017-18-GEN-10451/(SA-II).

\end{document}